\PassOptionsToPackage{hyphens}{url}
\documentclass[preprint2,times,tighten]{aastex6}
\pdfoutput=1 
\usepackage{url}
\usepackage{amsmath,amstext,natbib,float,epsfig,multirow,appendix}
\usepackage[figure,figure*]{hypcap}

\newcommand{\hMpc}{\ensuremath{h^{-1}\,\mathrm{Mpc}}}

\shorttitle{Filaments and MaNGA}
\shortauthors{Krolewski et al.}

\begin{document}

\bibpunct[]{(}{)}{;}{a}{}{;}

\title{Alignment between filaments and galaxy spins from the MaNGA integral-field survey}
\author{Alex Krolewski\altaffilmark{1,2}, Shirley Ho\altaffilmark{2,4,5,7}, Yen-Chi Chen\altaffilmark{3}, P.~F. Chan\altaffilmark{2,6}, Ananth Tenneti\altaffilmark{4}, Dmitry Bizyaev\altaffilmark{8,9}, Katarina Kraljic\altaffilmark{10}}
\email{krolewski@berkeley.edu}
\altaffiltext{1}{Department of Astronomy, University of California at Berkeley, New Campbell Hall, Berkeley, CA 94720, USA}
\altaffiltext{2}{Lawrence Berkeley National Lab, 1 Cyclotron Road, Berkeley, CA 94720, USA}
\altaffiltext{3}{Department of Statistics, University of Washington, Seattle, WA 98105, USA}
\altaffiltext{4}{Department of Physics, Carnegie Mellon University, Pittsburgh, PA 15213, USA}
\altaffiltext{5}{McWilliams Center for Cosmology, Carnegie Mellon University, Pittsburgh, PA 15213, USA}
\altaffiltext{6}{The Chinese University of Hong Kong, Shatin, NT, Hong Kong SAR, The People's Republic of China}

\altaffiltext{7}{Berkeley Center for Cosmological Physics, University of California at Berkeley, New Campbell Hall, Berkeley, CA 94720, USA}
\altaffiltext{8}{Apache Point Observatory and New Mexico State
University, P.O. Box 59, Sunspot, NM, 88349-0059, USA}
\altaffiltext{9}{Sternberg Astronomical Institute, Moscow State
University, Moscow, Russia}

\altaffiltext{10}{Institute for Astronomy, University of Edinburgh, Royal Observatory, Blackford Hill, Edinburgh, EH9 3HJ, UK}

\begin{abstract}

Halos and galaxies acquire their angular momentum during the collapse of surrounding large-scale structure.
This process imprints alignments between galaxy spins and nearby filaments and sheets.  
Low mass halos grow by accretion onto filaments, aligning their spins with the filaments,
whereas high mass halos grow by mergers along filaments, generating spins perpendicular
to the filament.
We search for this
alignment signal using filaments identified with the ``Cosmic Web Reconstruction'' algorithm
 applied to the Sloan Digital Sky Survey Main Galaxy Sample and 
galaxy spins from the MaNGA integral-field unit survey.  MaNGA produces a map of the galaxy's rotational velocity, allowing direct measurement
of the galaxy's spin direction, or unit
angular momentum vector projected onto the sky.
We find no evidence for alignment between galaxy spins and filament directions.
We do find hints of a mass-dependent alignment signal,
which is in 2-3$\sigma$ tension with the mass-dependent alignment signal in the MassiveBlack-II 
and Illustris hydrodynamical simulations.
However, the tension vanishes when galaxy spin is measured
using the H$\alpha$ emission line velocity rather than stellar velocity.
Finally, in simulations we find that the mass-dependent transition
from aligned to anti-aligned dark matter halo spins is not
necessarily present in stellar spins: we find a stellar spin transition
in Illustris but not in MassiveBlack-II, highlighting the sensitivity
of spin-filament alignments
to feedback prescriptions and subgrid physics.

\end{abstract}


\section{Introduction}
Dark matter protohalos acquire their angular momentum through tidal torquing by neighboring large scale structure \citep{peeb69,doro70,whi84}.
In the linear regime, angular momentum grows linearly with time and is aligned along the intermediate eigenvector of the tidal tensor
(i.e.\ the traceless part of the Hessian of the potential $\Phi$).
However, tidal torque theory is only qualitatively correct in the nonlinear regime, as nonlinear evolution significantly weakens
the spin alignment \citep{por02} and drives alignments with other preferred directions.  In the Zel'dovich picture of structure formation,
collapse occurs sequentially along the eigenvectors of the tidal tensor \citep{zel70}, forming anisotropic structures such as sheets (one direction
of collapse and two of expansion) and filaments (two directions of collapse and one of expansion).  Halos in filaments therefore
acquire spin parallel to the filament, as matter collapses and rotates in the plane perpendicular to the filament \citep{pich+11,codis+12}.
  N-body and hydrodynamic simulations have confirmed this result for low-mass halos ($M \lesssim 10^{12}$ M$_{\odot}$), while finding
that mergers align high-mass halo spins perpendicular to filaments
by converting motion along the filament into spin \citep{2005ApJ...627..647B,ac+07,hahn+07_evolution,codis+12,trow+13,ac14,dub+14,codis_IA,gan18,wang18,wang18b}.

Observations probe the spin of baryons within the galaxy rather than the spin of dark matter in the host halo.
Initially, the baryons and dark matter share the same angular momentum distribution and the
baryons conserve angular momentum as they collapse, creating a rotation-supported disk \citep{fe80,blum86,mo98}.
The size and profile of the baryonic disk, as computed from the angular momentum profile and dimensionless spin $\lambda$
of halos in N-body simulations, are roughly consistent with observations \citep{fall83,bull01}.
This simple picture cannot be correct in detail, however, since the baryons are subject to different physical processes than the dark matter,
including dissipation, disk instabilities, and feedback-driven outflows \citep{dan15}.  These processes lead to misalignments
between the spins of the dark matter and the baryons \citep{vdb02,bett12}.  As a result,
the mass-dependent alignment transition found
in simulations, which typically use gravity-only
$N$-body codes 
\citep[but see][ for spin-filament
alignments in hydrodynamic simulations]{dub+14,codis_IA,wang18b},
may not be present in observations or hydrodynamic simulations
of galaxy alignments.

Alignments between galaxy spins and large-scale structure have been measured using imaging to infer
the galaxy's inclination and spin axis
from its the axis ratio and position
angle.  At $z \sim 0$, studies have found suggestive but ultimately not
significant evidence for correlations between the chirality of neighboring galaxy spins \citep{slo+09,lee11,and+11}.
Studies of alignments between galaxy spins and large-scale structure have reached conflicting conclusions.
Early studies from small galaxy samples
in photographic plate surveys yielded
weak and conflicting results 
\citep[][ and references therein]{gregory+81,dekel+85,cab98}.
More recent results from larger samples
suggested
that spiral galaxies are aligned along the intermediate axis of the tidal tensor, in accord with predictions from tidal torque theory \citep{lp02,le07}, and are
therefore aligned perpendicular to filaments \citep{jon+10,zhang+15}.  However, a number of
studies within the past few years
have found little support for tidal torque theory predictions and instead suggest that low-mass spiral spins are parallel to filaments while higher mass elliptical or lenticular
spins are perpendicular to filaments \citep{temp13,tl13,pah+16}.

Alignments between galaxy spins are of particular interest as they are a major source of systematic error for weak lensing shear measurements,
particularly for upcoming missions such as LSST \citep{lsst_book}, WFIRST 
\citep{wfirst_paper} and EUCLID \citep{euclid_paper} that aim to measure the dark energy equation of state \citep{bridle+07,kirk+12}.
For disk galaxies, galaxy ellipticities arise from galaxy spins and are quadratic in the tidal field under tidal torque theory
\citep{lp00,cat+01}, while for elliptical galaxies, ellipticity arises directly from stretching by tidal fields and is linearly related to the tidal field \citep{hs04}.
 As a result, measurements
of alignments between galaxy spins and the surrounding tidal field or large-scale structures (clusters, filaments, sheets and voids)
can inform physical models of intrinsic alignments,
particularly for disk galaxies,
whose intrinsic alignment remains poorly constrained
\citep{hirata+07,mandelbaum+11}.

We measure spin-filament alignments using galaxy spins determined from integral-field kinematics rather than from
galaxy imaging.
Our method is complementary to imaging-based spin measurements, as it has very different sources of systematic error.
Galaxies often have low-surface brightness features
such as spiral arms or tidal tails, and therefore the galaxy shape may depend strongly on the measurement method, e.g.
which isophote is used
\citep[see Fig.\ 1 in][]{kirk_observation_review}.  Similarly, galactic bulges can bias shape measurements even for very late-type galaxies
\citep{and+11}.  While careful modelling including bulge/disk decomposition can alleviate this bias \citep[e.g.][]{temp13,tl13}, using
kinematics to measure galaxy spin eliminates the need for complex models of galaxy morphology and their associated uncertainty.

In this paper, we measure the alignment between filaments identified in the SDSS Main Galaxy Sample and galaxy spins
measured from MaNGA kinematics.  We use the filament catalog of \citet{chen+16}, which finds filaments as ridges in the density field
using the subspace-constrained mean-shift algorithm
(Section~\ref{sec:methods}).
We find no preference for spin-filament alignments in our overall sample of $\sim$2700 galaxies, and we validate our results
by finding similar alignments between galaxies and the Bisous
model filaments of \citet{temp+14}
(Section~\ref{sec:results}).  We compare our results to spin-filament alignments in hydrodynamical simulations
by measuring the mass-dependence of the alignment signal, and find 
2-3$\sigma$ tension when using spins
measured from the stellar continuum, but no tension
when using spins measured from the H$\alpha$ emission line (Section 4).  Finally, we compare our results to previous findings
 and conclude in Section~\ref{sec:conclusions}.
 
In this paper we use a flat $\Lambda$CDM cosmology
with $\Omega_m = 0.3$ and $h = 0.7$.  We convert
all masses to $M_{\odot}$ for inter-comparison
between observations and simulations.

\section{Methods and Data}
\label{sec:methods}

\subsection{Filament finder}
\label{sec:ff}

A variety of methods have been used to find filaments in observations and simulations, 
including approaches identifying filaments as
eigenvectors of the deformation tensor 
\citep{hahn+07_evolution,jas+10}, velocity shear tensor \citep{libe+13}, or Hessian of the density field \citep{ac07};
identification of filaments as
ridges in the density field \citep{sous+08,chen+15_methods}; 
and searches for cylindrical arrangements of galaxies \citep{temp+14}.  For a comprehensive overview, see \citet{caut14}.

We use the publicly available \textit{Cosmic Web Reconstruction} filament algorithm\footnote{\url{https://sites.google.com/site/yenchicr/}}
\citep{chen+16}
to identify filaments in the SDSS Main Galaxy Sample.  
This filament finder identifies filaments as curves
in two-dimensional ($\alpha$, $\delta$) slices of width
$\delta z = 0.005 \sim 20$ Mpc.  This yields a well-defined orientation for every point
on the filament and makes it easy to cross-correlate
with the spin
of nearby galaxies.
The filament finder is explained in detail in \citet{chen+15_methods}, so we only provide
a brief description here.  
Our filament catalog differs slightly from the publicly available catalog of \citet{chen+15_methods},
as it extends to lower redshift and uses slightly different thresholding to remove noisy filaments.

The filament finder operates on a smoothed density field
created from the positions of galaxies in the SDSS Main Galaxy Sample \citep{bla+05}
and the LOWZ and CMASS samples from BOSS \citep{ala+15}, with a redshift-dependent Gaussian smoothing
kernel that ranges between 5 and 10 Mpc \citep[Fig.\ 6 in][]{chen+16}.
It identifies filaments as density ridges of the smoothed density field, or local
maxima along the second eigenvector of the Hessian of the density field.  

\begin{figure*}
\includegraphics[width=\textwidth,clip=true,trim=60 20 60 30]{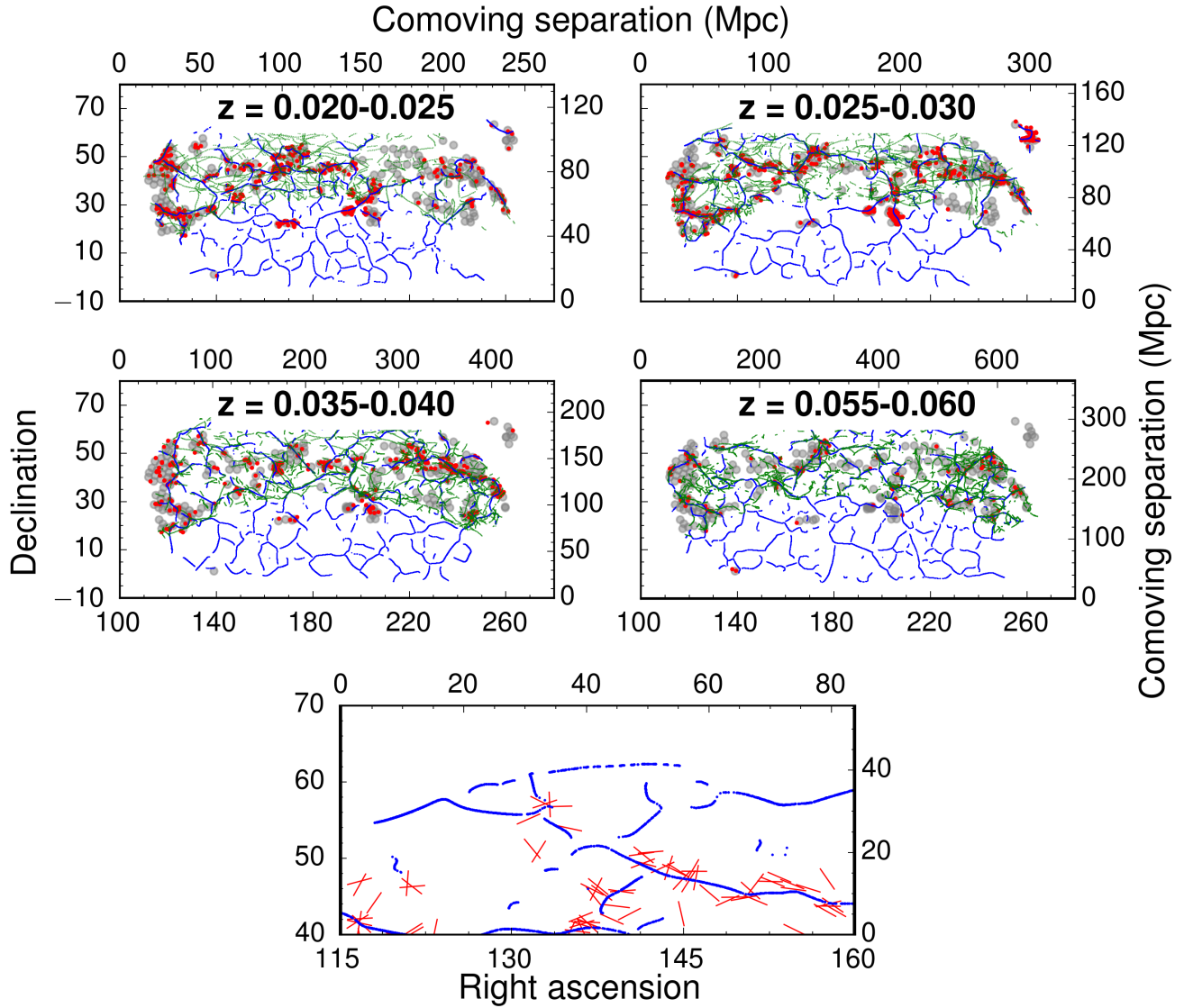}
\caption[]{\small \textit{Upper four panels:} MaNGA galaxies (red) located within 40 Mpc of \textit{Cosmic Web Reconstruction} filaments (blue) and \citet{temp+14} Bisous filaments (green).  Slices were chosen
as the lowest redshift slice with filaments ($z = 0.02$) and the three quartiles of the MaNGA galaxy
redshift distribution ($z = 0.027$, 0.036, and 0.052).  Gray circles indicate
MaNGA plates released with MPL-6.
\textit{Lower panel:}  Comparison between galaxy spins (red lines) and \textit{Cosmic Web Reconstruction} filaments (blue lines) for a section of the sky at $z = 0.025-0.03$.
}
\label{fig:filaments_galaxies}
\end{figure*}

The filament
finder uses two-dimensional slices of width $\delta z = 0.05$ ($c \delta z = 1500$ km s$^{-1}$ $\sim20$ Mpc); in each slice, it finds filaments in an equirectangular projection of equatorial coordinates ($\alpha$, $\delta$)
using only galaxies in the North Galactic Cap (Figure~\ref{fig:filaments_galaxies}).  We find filaments between $z = 0.02$ and $z = 0.15$,
with the lower limit set by the sparsity of SDSS galaxies at $z < 0.02$ and the upper limit
set by the maximum redshift of MaNGA galaxies ($z=0.15$).
At these redshifts the filament finder primarily
uses galaxies from the Main Galaxy Sample.
 We eliminate galaxies in the 10\% least dense environments, defined
using the distance to the 30th-nearest neighbor. This eliminates noisy filaments from very low-density regions without removing too many filaments.
Varying the thresholding criteria does not qualitatively change the results in Table~\ref{tab:alignments_by_category}.

We define the filament orientation at each
point as the first principal component of the covariance matrix of the positions of the ten nearest neighbor points.
We estimate the uncertainty on the filament directions at each point by measuring the local
filament orientation for 100 bootstrap resamples of the filament catalog.

Filaments are identified in 2D rather than 3D for ease of interpretation: collapsing along the line of sight eliminates spurious filaments created by redshift-space
distortions and allows us to better model the strong redshift dependence of galaxy density, which requires
a redshift-dependent smoothing length
\citep{chen+16}.  Furthermore, previous work measuring three-dimensional spin-filament alignments
 has found that line of sight biases in both galaxy spins and filaments
creates strong spurious alignment signals which must be corrected \citep{temp13,tl13}.
From simulations, we expect that using 2D rather than 3D filaments reduces our signal by $\sim 40\%$ (Appendix~\ref{sec:3d_vs_2d}); thus we believe the moderate loss in signal is worth the substantial
reduction in systematic errors.

In Figure~\ref{fig:filaments_galaxies} we plot the MaNGA galaxy sample
(with $z > 0.02$ and distance to filament $d_F < 40$ Mpc) and the \textit{Cosmic Web Reconstruction} filaments in four redshift slices: $z = 0.02-0.025$, the lowest-redshift slice, and the slices
containing the three quartiles of the redshift distribution, $z = 0.025-0.03$, $0.035-0.04$ and $0.055-0.06$.
In Figure~\ref{fig:filaments_galaxies}, most galaxies are clearly closest
to a single filament,
indicating that
confusion between filaments will not contribute significantly to noise in the measurement.

To check our results,
we measure alignments with the \citet{temp+14} filament catalog\footnote{
Available on Vizier, \url{http://vizier.cfa.harvard.edu/viz-bin/VizieR?-source=J/MNRAS/438/3465}, including a filament
catalog; catalog of filament points;
and catalog of all galaxies used to construct
the filament catalog and their velocity-corrected distances.}, which was also derived from galaxies in the SDSS Main Galaxy Sample, with
$0.009 < z < 0.155$. \citet{temp+14} use a very
different method
from \citet{chen+15_methods}: they find filaments using the Bisous model,
a marked point process model which fits the galaxy
data to 
a filamentary network composed of connected
cylinders of fixed width.  They find filaments
in three dimensions, suppressing
peculiar velocities by estimating the
velocity dispersions for galaxy groups.
We measure alignments using galaxies within
20 $h^{-1}$ Mpc of filaments and with a velocity-corrected
distance from the \citet{temp+14} catalog,
yielding a sample of 3028 galaxies.
For each galaxy, we consider its alignment
with the plane-of-sky projection of the nearest
\citet{temp+14} filament.
We compare the \citet{temp+14} and \citet{chen+15_methods}
filaments in Figure~\ref{fig:filaments_galaxies}; 
\citet{temp+14} identify significantly smaller-scale filaments,
but on larger scales both methods recover similar filaments.
Despite the substantial methodological
differences between the two filament finders,
we find largely similar
alignments (Section~\ref{sec:results}).

\subsection{MaNGA galaxies}
\label{sec:manga}

Mapping Nearby Galaxies at Apache Point Observatory (MaNGA) is an integral-field survey that aims to obtain spectra of 10,000
nearby galaxies \citep{bun+15}.
It began in July 2014 as part of SDSS-IV \citep{blanton:2017} and is planned to continue until 2020.  MaNGA uses the 2.5-m SDSS telescope at Apache
Point Observatory in New Mexico \citep{gunn+06}
and the dual fiber-fed BOSS spectrographs \citep{smee+13}, but rather than allocating a single fiber per galaxy like previous SDSS surveys,
each plate contains 17 pluggable Integral Field Units, each of which consists of hexagonal bundles containing between 19 and 127 fibers with
typical spatial resolution of 2.5'' or 1.8 kpc at $z = 0.03$ \citep{dro+15}.
The dual spectrograph design enables a wavelength coverage of 3600--10000 \AA\ with a velocity resolution of 70 km s$^{-1}$ \citep{smee+13}.
Typical exposure times of 3 hours ensured S/N of 5 at the outskirts of targeted galaxies, and much greater towards the center \citep{law_obs_strat}.
Spectrophotometric calibration is accurate to $<5 \%$ \citep{yan_specphot} and the data reduction pipeline is described in \citet{law_drp}.

\begin{figure}
\hspace{-10pt}
\includegraphics[width=0.5\textwidth,clip=true,trim=40 50 0 0]{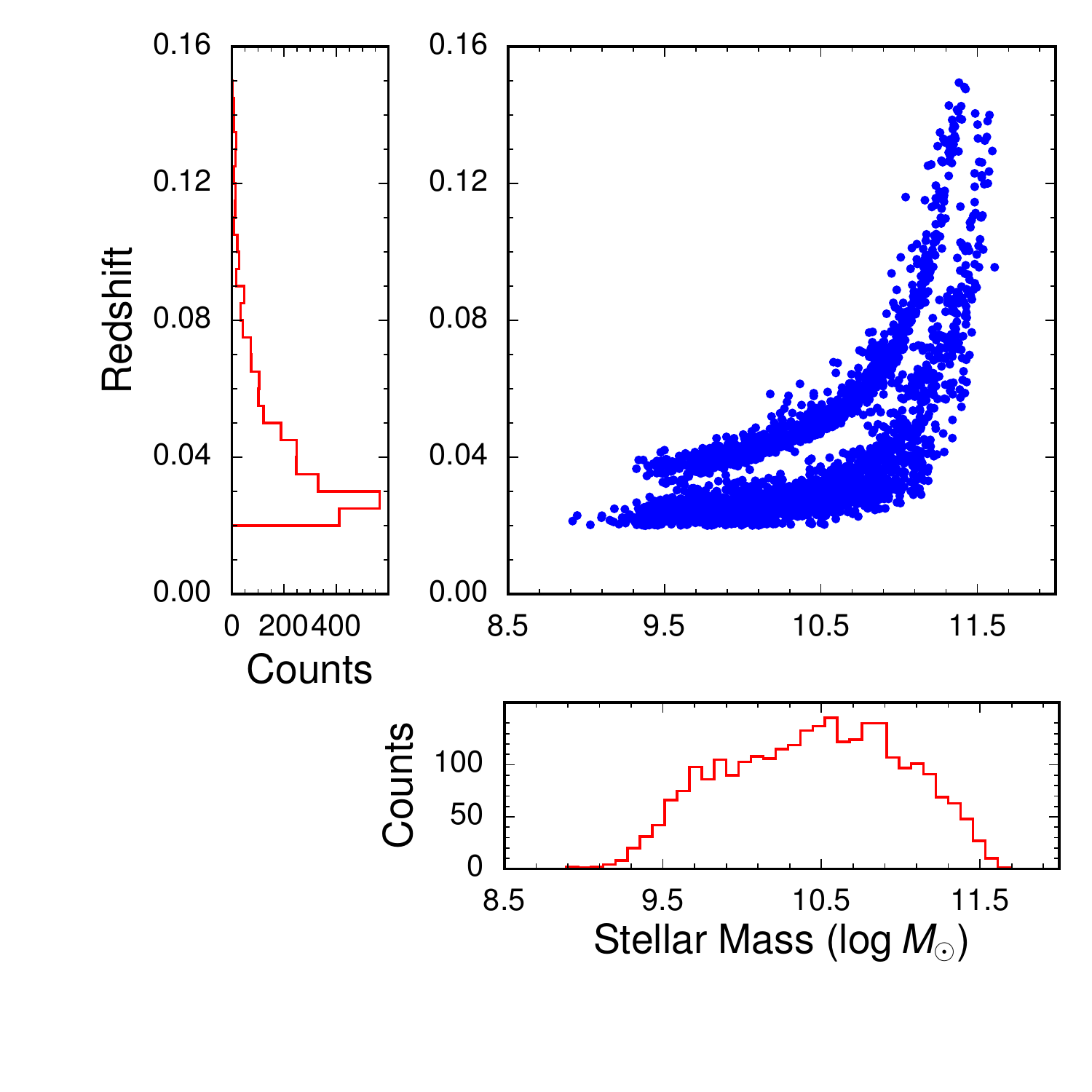}

\caption[]{\small Redshift (top) and stellar mass (bottom) distributions of the MaNGA sample of 2736 galaxies (selected according
to criteria in Section~\ref{sec:spins}).
}
\label{fig:redshift_mass_dist}
\end{figure}

The MaNGA targeting sample consists of 10,000 galaxies with $0.01 < z < 0.15$ (median $z \sim 0.03$).  The sample was chosen to have a flat
number density distribution in absolute $i$-band magnitude $M_i$ while maximizing the spatial resolution  and ensuring IFU coverage to a few times the half-light radius $R_e$ \citep{yan_sample}.  
As a result, stellar mass is highly correlated with redshift for the MaNGA sample, since galaxies 
of a given mass (and thus radius) are preferentially targeted at a redshift where the IFUs cover a few $R_e$ (Figure~\ref{fig:redshift_mass_dist}).
The double-peaked redshift-mass distribution is a result
of the two-tiered MaNGA selection process, consisting of the Primary sample with
coverage to $1.5R_e$ and the Secondary sample with coverage to $2.5R_e$.
Galaxies are assigned to plates
via a tiling algorithm that is unbiased with respect to environment, and to IFUs in a way that maximizes the number of galaxies covered to the
appropriate radius (1.5 $R_e$ for Primary sample and 2.5 $R_e$ for the Secondary sample).

We use the MPL-6 data release of MaNGA with v2\_3\_1 of the Data Reduction Pipeline and v2.1.3 of the Data Analysis Pipeline.
MPL-6 contains 4687 galaxy data cubes observed
between March 2014 and July 2017, of which 70 are repeat observations. We subsequently reduce
our sample to 2736 galaxies via a variety of quality cuts.
We remove 85 galaxies with the 
the \textsc{CRITICAL} \textsc{DRP3QUAL}
maskbit set, which indicates a variety of problems ranging
from unmasked cosmic rays to IFUs partially falling
out of the plate;
426 galaxies targeted as part of ancillary programs, which
lack well-defined selection weights;
393 galaxies with $z < 0.02$; and
478 galaxies lying beyond the 40 Mpc
radius of influence for galaxy-filament
alignments found in \citet{chen+18}.
Finally, we remove galaxies with 
poorly measured spins (see Section~\ref{sec:spins}): 19 galaxies
lacking a sufficient number of points
to fit a spin; 170 galaxies with multiple
galaxies inside the IFU (Figure~\ref{fig:example_multiple_gals});
and 858 galaxies with position angle
error $>5^{\circ}$, which we find
by visual inspection
to generally have poorly-defined
spins.

We weight each galaxy to create a volume-limited
sample \citep{wake+17}
that is appropriate to compare to simulations.
Specifically, we weight each galaxy
by the ``esrweights''
\citep[Equation A12 in][]{wake+17},
the effective volume over which it could have been observed.
The weights are necessary because MaNGA is not a volume-limited sample; the flat distribution in $M_i$ leads to to biases towards higher
luminosity at fixed mass, biasing galaxy colors and
inclinations
\citep{wake+17}.
All results in Sections~\ref{sec:results}
and 4
use weighted mean dot products
and bootstrap resampling to compute the standard error of the weighted means.

The MaNGA sample
is complete to $\log{(M_{\star}/M_{\odot})} = 9.61$
for the Secondary sample and $\log{(M_{\star}/M_{\odot})} = 9.10$ for the Primary sample \citep{wake+17}; thus, we require $\log{(M_{\star}/M_{\odot})} > 9.6$
for comparison to mass-dependent alignment in simulations
(Section 4), limiting this comparison to
2551 galaxies.
Additionally, the Secondary sample is incomplete for
highly
inclined galaxies slightly above $\log{(M_{\star}/M_{\odot})} = 9.61$
\citep{wake+17}, although such galaxies
only constitute a small portion
of the sample in the lowest mass bin.

Gross galaxy properties such as absolute magnitude, color, stellar mass, and photometric shape are extracted from the MaNGA targeting catalog, v1\_0\_1 of the NASA-Sloan Atlas \citep{bla+11}.  This catalog is superior to the SDSS catalog for photometry of bright extended galaxies.  We use magnitudes and stellar masses
from elliptical Petrosian photometry,
recommended as the most reliable photometry in the catalog\footnote{See \url{http://www.sdss.org/dr13/manga/manga-target-selection/nsa/}}.
We use the Sersic photometry for axis ratios and photometric position angles.
The stellar masses and star formation rates are calculated using the \textsc{k-correct} code \citep{bla+03} with a \citet{chab+03} initial mass function.
In Figure~\ref{fig:redshift_mass_dist}, we show the redshift and mass distribution of the final sample of 2736 galaxies.

We use Galaxy Zoo  for morphological classification \citep{gz+11},
matching each MaNGA galaxy to the nearest Galaxy Zoo source within 0.5''.  The Galaxy Zoo catalog gives a probability that each galaxy is a spiral (clockwise or counter-clockwise),
elliptical, edge-on, merger or unknown morphology.
To study the morphological dependence of spin-filament alignments, we only use galaxies with
a $> 50$\% probability of any single
classification.  
Edge-on galaxies are defined as galaxies with axis ratio $r < 0.3$ rather than using Galaxy Zoo, since Galaxy Zoo classifies any galaxy
with spiral structure as a spiral even if it is nearly edge-on. 
Our final sample therefore contains 1039
elliptical galaxies, 676 spiral
galaxies, and 344 edge-on
galaxies, with the rest unclassified.

We use the stellar velocity maps produced by the Data Analysis Pipeline (DAP) for MPL-6 (Westfall et al., in prep),
which uses the penalized-pixel fitting method (pPXF) \citep{cap+04} to determine kinematic parameters.
For the spectrum in each spaxel, the DAP first fits the stellar continuum using the MILES stellar library
\citep{miles:2006} and masking emission lines.
Emission lines are subsequently fit, fixing the stellar continuum to the previously-determined best-fit values.

To check the robustness of our results, we measure galaxy spins
from both the stellar continuum and 
the H$\alpha$ emission line velocity maps.
We apply the same fitting methods (Section~\ref{sec:spins})
to both velocity maps.
These measurements trace different physical components
of the galaxy: the stellar
continuum traces the stars while the emission
line traces the gas. 

\subsection{Galaxy spins}
\label{sec:spins}

\begin{figure}
\includegraphics[width=0.5\textwidth]{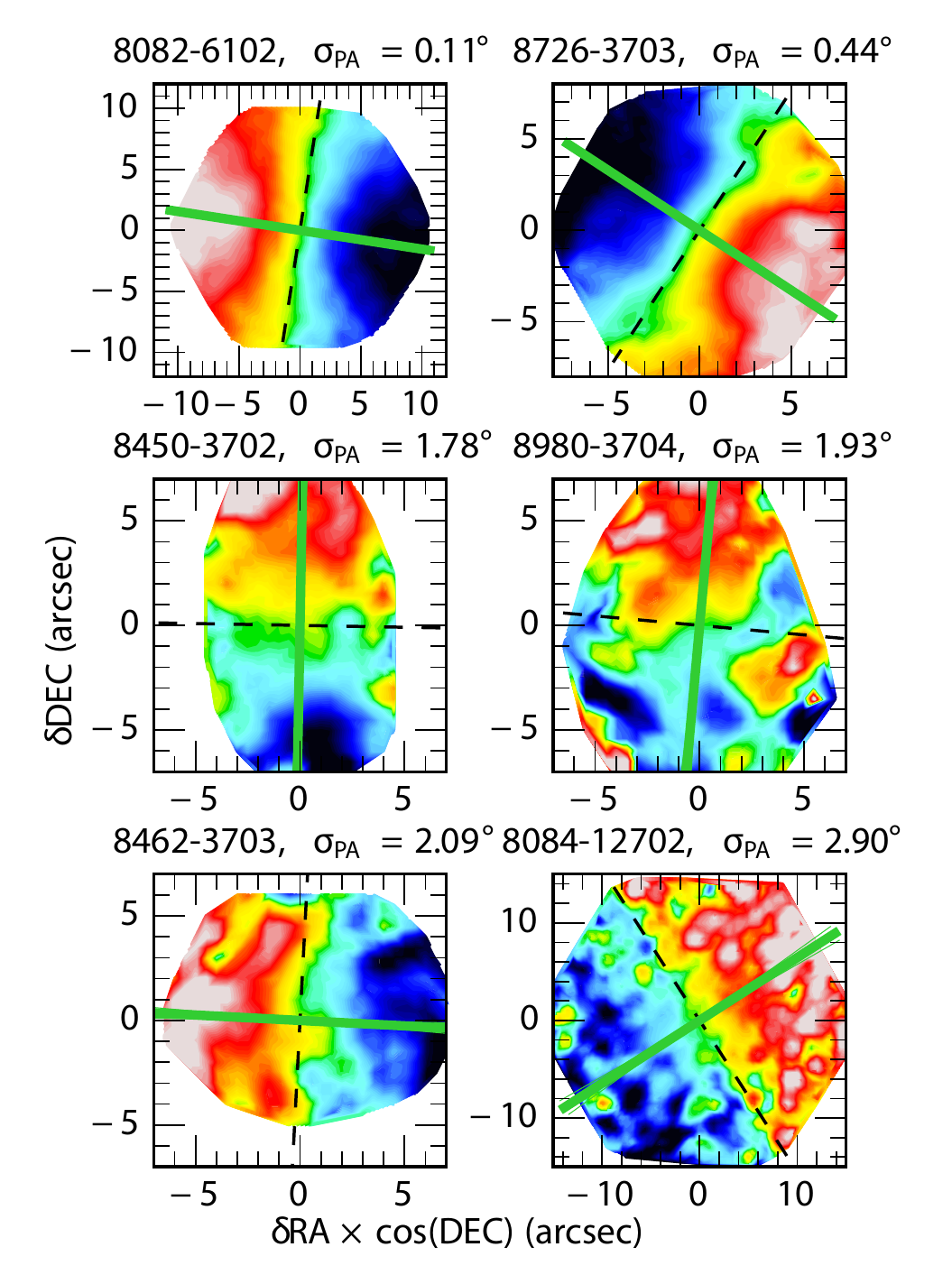}
\caption[]{\small 6 randomly selected galaxies with stellar velocity maps and \textsc{FIT\_KINEMATIC\_PA} fits.  
The best-fit kinematic position angle is the thick green line and the spin vector is the dashed black line.
The thin green lines (often obscured by the thick green line) show the 1-$\sigma$ uncertainty on the position angle.
The title gives the plate and IFU ID uniquely identifying
each observation and the error on the PA in degrees.
}
\label{fig:example_fits}
\end{figure}

We determine the spin vector for each galaxy by measuring the kinematic position angle
using integral-field data from the MaNGA survey.  Specifically, the plane-of-sky projection of the spin vector
is perpendicular to the kinematic position angle (Figure~\ref{fig:example_fits}).
For each galaxy we determine a single
global position angle (and thus spin direction) from the full datacube.  We apply
the \textsc{FIT\_KINEMATIC\_PA} routine \citep{kraj+06} to determine the kinematic position angle for each galaxy
from the stellar velocity maps, using velocities from the unbinned spaxels (see Appendix~\ref{sec:spin_details} for further
details).
Our method is necessarily two-dimensional,
consistent with our two-dimensional filament finder. In accordance with the two-dimensional nature of our measurement,
hereafter we refer to the plane-of-sky projection of the spin as the galaxy spin vector.
While the three-dimensional spin
could be estimated using the galaxy's axis ratio to find the inclination \citep{hg84},
this method requires an estimate of the galaxy's intrinsic thickness;
assumes that the galaxy's shape can be approximated by an oblate spheroid, which may not be valid
for elliptical galaxies; and could be biased by the isophote used or the presence of a galactic
bulge \citep{and+11,kirk_observation_review}.  Moreover, estimating the three-dimensional spin
from the galaxy's shape necessarily leads to anisotropic errors between the plane of sky and the line of sight
and potentially an inhomogeneous distribution of inclinations \cite[e.g.][]{temp13,tl13}.

\begin{figure}
\includegraphics[width=0.5\textwidth,clip=true,trim=10 0 15 0]{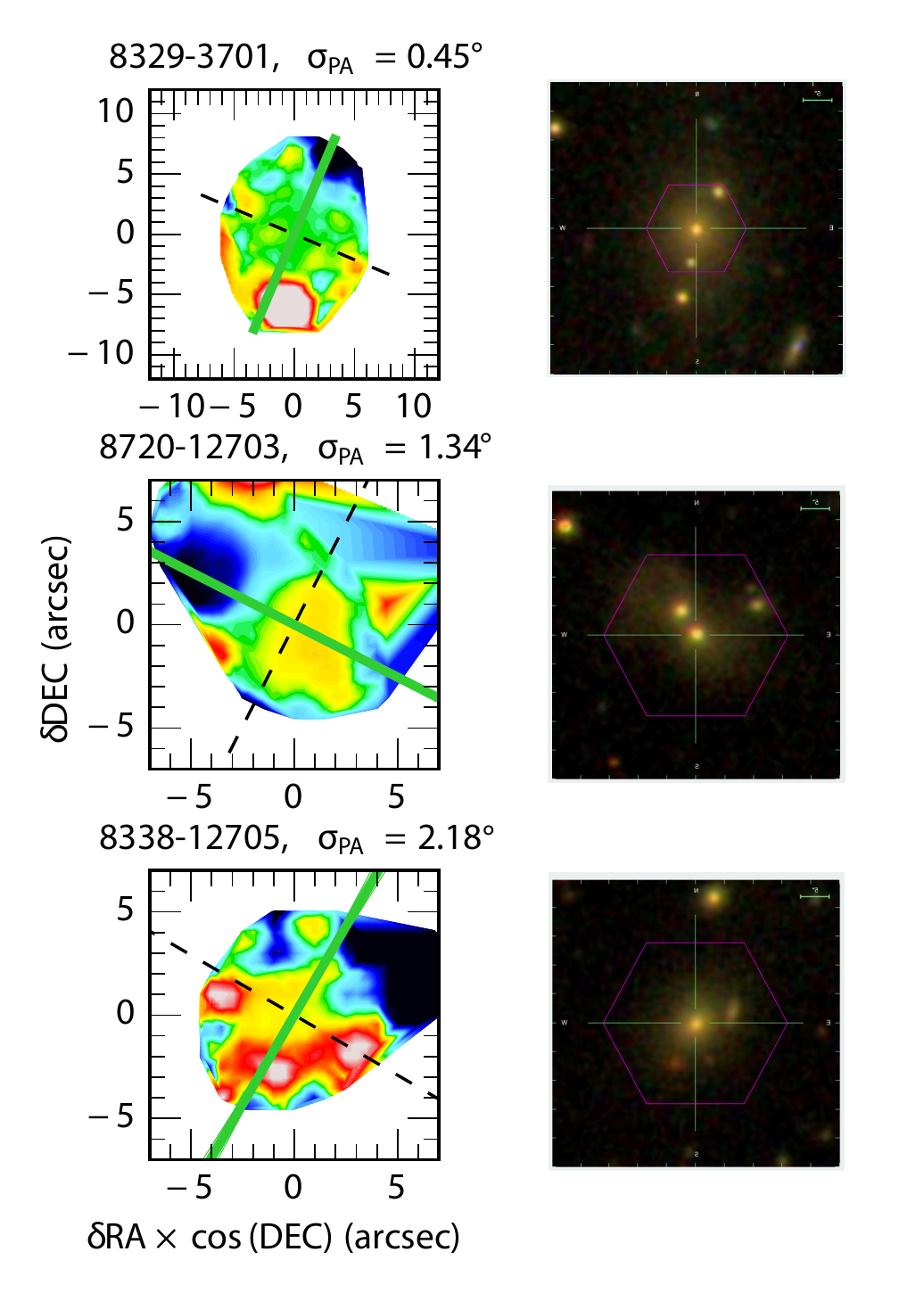}
\caption[]{\small 3 randomly selected cases with multiple galaxies inside the IFU. 
Panels on the left show the stellar
velocity maps and \textsc{FIT\_KINEMATIC\_PA} fits.  Panels on the right show the SDSS image with the MaNGA
IFU overlaid.
Although the fitting errors on these galaxies are formally smaller
than the cutoff for poorly measured spins (error $> 5^{\circ}$),
it is clear from comparison to the SDSS images
that the fit is spurious due to multiple galaxies in the IFU.
}
\label{fig:example_multiple_gals}
\end{figure}

We show 6 randomly selected fits in Figure~\ref{fig:example_fits}.  The output of \textsc{FIT\_KINEMATIC\_PA}
agrees well with the position angle one would identify by eye.  However, \textsc{FIT\_KINEMATIC\_PA} fails in cases where there
are multiple kinematically-distinct galaxies in the IFU.  In these cases, \textsc{FIT\_KINEMATIC\_PA} spuriously identifies the line connecting
the galaxies as the position angle (Figure~\ref{fig:example_multiple_gals}).  We identify these cases by searching for galaxies with multiple SDSS $r < 20$ sources located
within the IFU and visually inspect each image to distinguish contaminants from foreground stars, background galaxies, and errors
in SDSS photometry.  We find and exclude 171 galaxies with spurious fits due to multiple kinematically-distinct galaxies in the IFU.

Visual inspection shows that the velocity maps become increasingly noisy, with poorly
defined rotation, when $\sigma_{\textrm{PA}} > 5^{\circ}$.  As a result, we remove these 858 galaxies from our measurement.
Using a stricter cut of $\sigma_{\textrm{PA}} < 3^{\circ}$
changes the results presented in Table~\ref{tab:alignments_by_category}
by $\lesssim 1\sigma$.

In the bottom panel of Figure~\ref{fig:filaments_galaxies}, we plot galaxy spin vectors and filaments for a small region of the sky at $z = 0.025-0.03$ to
illustrate the alignment measurement.  We are searching for a weak alignment identifiable statistically but not visually.

\begin{figure}
\includegraphics[width=0.5\textwidth,clip=true,trim=15 0 0 0]{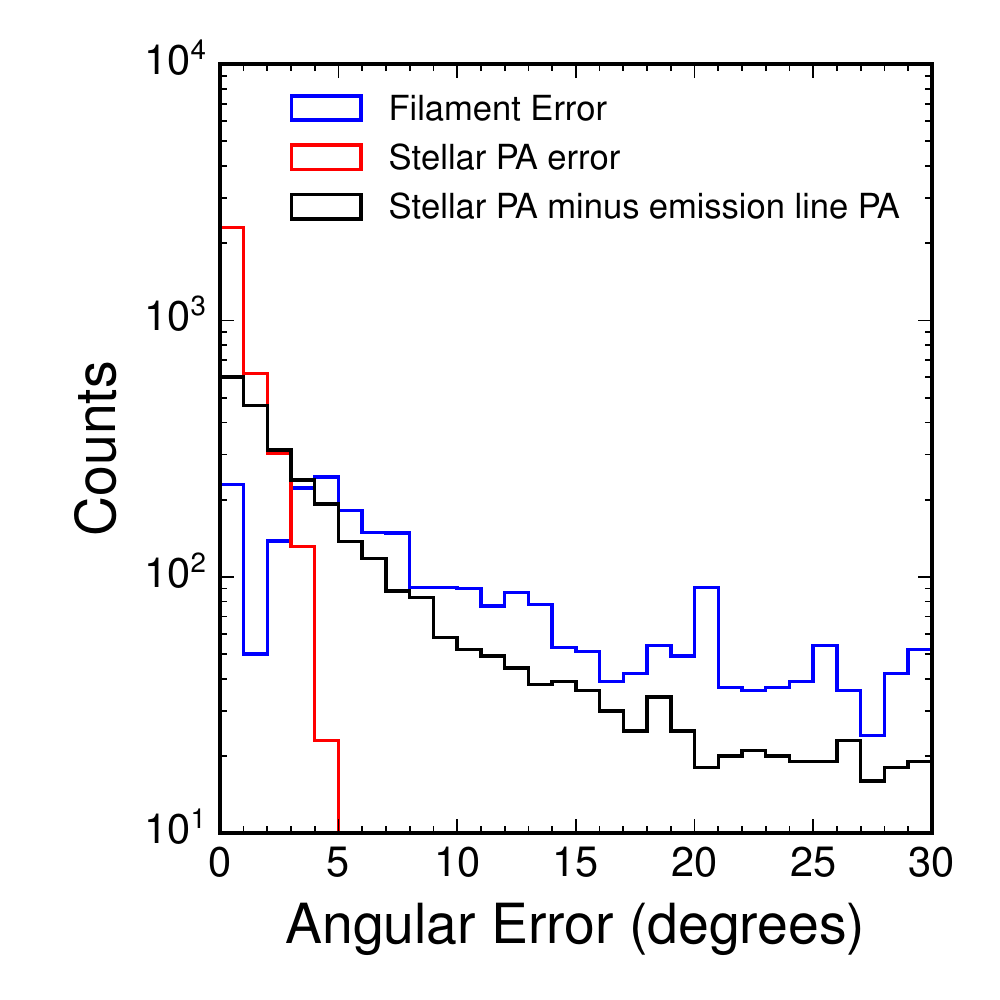}
\caption[]{\small Comparison between the error on the filament angle
nearest to each galaxy (blue), the error on the stellar position angle (red),
and the dispersion between the stellar position angle and the emission line
position angle (black).
}
\label{fig:fils_vs_pa_err}
\end{figure}

In Figure~\ref{fig:fils_vs_pa_err}, we show that the measurement errors on the filaments
(blue) dominate the errors
on the galaxy position angles (red).
We also plot the distribution of stellar minus emission
line position angle; since this dispersion is the quadrature sum of the measurement error on the position angle and the true dispersion between the stellar
and emission line spins, it provides an upper bound
on the position angle error.  This dispersion
is still smaller than the filament error,
showing that the filament error must be greater than the
position angle error.

\subsection{Mock spins and filament catalogs from hydrodynamical simulations}
\label{sec:mbii}

We compare our results to galaxy alignments measured in two 
publicly available
cosmological hydrodynamical
simulations, MassiveBlack-II\footnote{\url{http://mbii.phys.cmu.edu/}} \citep{khan+15} and Illustris-1\footnote{\url{http://www.illustris-project.org/data/}} \citep{vogels:2014,nelson:2015}.  Since
the spin-filament alignment signal is quite subtle, we require large box hydrodynamical
simulations ($L \gtrsim 100$ Mpc).

MassiveBlack-II is a cosmological simulation run using the smoothed particle hydrodynamics code $\textsc{GADGET}$
in a 100 \hMpc{} box with $\Omega_m = 0.275$, $h = 0.704$, and $\sigma_8  = 0.816$ \citep{khan+15}.  The simulation contains $2 \times 1792^3$ particles, with dark matter particle mass of $1.1 \times 10^7 h^{-1} M_{\odot}$
and gas particle mass $2.2 \times 10^6 h^{-1} M_{\odot}$.  MassiveBlack-II includes subgrid models for star formation and black hole feedback.
Star formation occurs according to the multiphase model of \citet{sh+03}, 
and young stars and supernovae provide feedback by heating the 
Gas is accreted onto black holes following Bondi accretion, limited to twice the Eddington accretion rate, and 5\% of the energy radiated by the accreting
black hole is deposited as feedback.
Halos are identified
using a friends-of-friends algorithm with linking length $b = 0.2$, and subhalos are identified using $\textsc{SUBFIND}$.  Halos and subhalos are required to have at least 40 dark matter and gas particles;
therefore, the stellar mass limit of the simulation is $1.26 \times 10^8 M_{\odot}$.

In MassiveBlack-II, the spin for each galaxy is defined as the unit stellar angular momentum vector.
Galaxy spins are only computed for subhalos with at least 1000 dark matter and star particles.
This corresponds to a stellar mass limit of $\log{(M_{\star}/M_{\odot})} = 9.5$ for the sample
with measured spins; thus, the spin subsample is complete for the mass range of the MaNGA sample
($\log{(M_{\star}/M_{\odot})} > 9.6$), confirming
the validity of comparing data to simulations over this mass range.
We also consider alignments between
filaments and gas spins in MassiveBlack-II,
where gas spins are computed for subhalos with
at least 1000 gas particles.

Illustris-1 is run using the moving-mesh code AREPO in a 75 \hMpc{} box with $1820^3$
gas and dark matter particles each for a dark matter particle mass of $6.3 \times 10^6 M_{\odot}$
and a gas particle mass of $1.3 \times 10^6 M_{\odot}$.   Subhalos are required to have
at least 20 particles\footnote{\url{http://www.illustris-project.org/data/docs/faq/\#cat3}};
therefore, the stellar mass limit is $2.6 \times 10^7 M_{\odot}$.
The cosmological parameters are $\Omega_m = 0.2726$, $h = 0.704$, and $\sigma_8 = 0.809$.
The subgrid physics is described
extensively in \citet{vogels:2013} and is similar to the subgrid physics in MassiveBlack-II,
but somewhat more elaborate: Illustris-1 uses variable wind speeds and mass loading
in the \citet{sh+03} galactic wind model, and Illustris-1 includes radio-mode AGN feedback
as well as quasar-mode feedback.

\begin{deluxetable*}{lccccc}
\tablecolumns{6}
\tablecaption{\label{tab:alignments_by_category} \textit{Cosmic Web Reconstruction} alignments for different subsamples}
\tablehead{
Sample & N & $\langle \cos{\theta} \rangle$ & SE  & Shuffle mean & $\sigma$ from shuffle
}
\startdata
All & 2736 & 0.6452 & 0.0075 & 0.6406 & 0.61 \\
$D_F < 0.6$ Mpc & 684 & 0.6474 & 0.0144 & 0.6438 & 0.25  \\
$0.6 < D_F < 1.4$  Mpc & 684 & 0.6532 & 0.0148 & 0.6427 & 0.71  \\
$1.4 < D_F < 3.0$ Mpc  & 684 & 0.6497 & 0.0146 & 0.6370 & 0.87  \\
$D_F > 3.0$ Mpc & 684 & 0.6257 & 0.0154 & 0.6365 & -0.70  \\
$M_{\star} < 10.02$  & 684 & 0.6601 & 0.013 & 0.6406 & 1.50  \\
$10.02 < M_{\star} < 10.47$ & 684 & 0.6288 & 0.0120 & 0.6425 & -1.14  \\
$10.47 < M_{\star} < 10.87$ & 684 & 0.6350 & 0.0125 & 0.6407 & -0.46 \\
$M_{\star} > 10.87$ & 684 & 0.6500 & 0.0147 & 0.6358 & 0.97 \\
$u-r < 1.70$ & 684 & 0.6562 & 0.0139 & 0.6417 & 1.04  \\
$1.70 < u-r < 2.09$& 684 & 0.6381 & 0.0149 & 0.6409 & -0.19  \\
$2.09 < u-r < 2.35$ & 684 & 0.6630 & 0.0130 & 0.6354 & 2.12  \\
$u-r > 2.35$ & 684 & 0.6072 & 0.0168 & 0.6373 & -1.79 \\
elliptical & 1039 & 0.6408 & 0.0130 & 0.6392 & 0.12 \\
spiral & 676 & 0.6423 & 0.0147 & 0.6399 & 0.16   \\
edge-on & 344 & 0.6239 & 0.0186 & 0.6392 & -0.82  \\
\enddata
\tablecomments{MaNGA spin-\textit{Cosmic Web Reconstruction} filament alignments for the entire sample and sub-samples split by distance
to filament ($D_F$), stellar mass (in units of $\log{ M_{\odot}}$), $u-r$ color, and morphology.  $\langle \cos{\theta} \rangle$ is the mean dot product
between the unit spin vector and the unit filament vector.  SE is the standard error of the mean, calculated using 50000 bootstrap resamples of the data.  We measure the expectation for random alignments
using 50000 shuffles of the data, and compute $\sigma$, the deviation between the data and the randoms in units of the standard error.}
\end{deluxetable*}

Halos in Illustris-1 are identified using a friends-of-friends algorithm with linking
length $b = 0.2$ on the dark matter particles, and subhalos are subsequently identified
using \textsc{SUBFIND}.  
For both MassiveBlack-II and Illustris, the masses quoted in this paper (both dark matter
and stellar) are defined as the total mass of all particles bound to a given \textsc{SUBFIND}
halo.
As in MassiveBlack-II, subhalo spin is defined
as the unit stellar angular momentum vector \citep{zjupa:2016}, summing over
all star particles within twice the stellar half-mass radius.

Angular momenta
are only calculated for subhalos with more than 300 dark matter particles,
yielding a stellar mass limit of $3.8 \times 10^8 M_{\odot}$ for halos
with the cosmic baryon fraction.  Therefore, as for MassiveBlack-II, 
it is valid to compare simulation and data alignments for MaNGA galaxies with $\log{(M_{\star}/M_{\odot})} > 9.6$.

Since we measure filaments using \textit{Cosmic Web Reconstruction} in two dimensions, some filaments
in our catalogs may just be cuts through sheets lying perpendicular to the plane of the sky.  Since halo alignments with sheets may be different
from halo alignments with filaments \citep[e.g.][ find a mass-dependent transition from alignment to anti-alignment with filaments, but mass-independent alignment between halos and sheets]{ac07}, our alignment
measurements are not directly comparable to three-dimensional filament alignment measurements in simulations.  Therefore, we compare our measurement
to mock observations in MassiveBlack-II and Illustris-1 reproducing the two-dimensional filaments used in the
observational work.

For both Illustris-I and MassiveBlack-II, we create a mock filament catalog for each of the 26 \textit{Cosmic Web Reconstruction} $\Delta z = 0.005$ slices between $z = 0.02$ and $z = 0.15$.  
These 26 filament catalogs allow us
to create a mock galaxy-filament
alignment measurement by matching
the redshift distribution of the MaNGA galaxies.
For each redshift slice, we
select subhalos in descending order of mass to match the number density of SDSS galaxies in that redshift slice.
We define filaments using subhalos rather than halos or dark matter particles because subhalos
are generally taken as proxies for galaxies in e.g.\ comparisons to the galaxy stellar mass function
\citep{khandai_massiveblack,vogels:2014}.
Since the completeness of the SDSS Main Galaxy Sample is $>90\%$
\citep{strau+02}, this procedure
yields a mock sample
representative of MGS.  We move the subhalos into redshift space 
and divide the box into 7 slices along the $z$ axis (width $14 \hMpc = 20$ Mpc), finding
filaments in two dimensions in each slice following the same method as in the data.  We generate the smoothed density field from
subhalos in the 90\% densest environments, match the smoothing bandwidth at each redshift to the bandwidth used in the data, and identify
the filament direction using local PCA.  We ignore the periodic boundary conditions of the box
when finding filaments.
With these 26 filament catalogs we can then make a mock observation of galaxy-filament alignment by randomly assigning each galaxy in the simulation to one of the 26 catalogs following the redshift distribution of the data
(see Section 4 for further details).

\section{Galaxy-filament alignments of entire sample}
\label{sec:results}

After the quality cuts described above and the redshift cut ($0.02 < z < 0.15$), we measure alignments with a sample
of 2736 galaxies.  We measure alignment using the mean dot product between the unit filament vectors and the unit galaxy
spin vectors.  A dot product of 1 indicates perfect alignment, 0 indicates perfect anti-alignment, and $2/\pi = 0.6366$ (i.e. the average value of $\cos{\theta}$ over the range 0 to $\pi$) indicates random alignment.  All
mean dot products are defined
as weighted means using the MaNGA
weights defined to recover
a volume-limited sample \citep{wake+17}.
Error bars are defined for the
weighted means using 50000 bootstrap
resamples of each galaxy subsample.
We compare the measured alignment
to a random signal generated from 50000
shuffles of the galaxy and filament catalogs;
if there are anisotropies
in the galaxy and filament catalogs,
the expectation for random alignments
will deviate from $2/\pi$.  In fact,
deviations from $2/\pi$ are modest
for all subsamples.

We find no evidence for alignments between galaxy spins and filaments, with a mean dot product of $0.6452 \pm 0.0075$,
an $0.61\sigma$ deviation from the shuffled dot product of $0.6406$.

In Table~\ref{tab:alignments_by_category},
we split the sample in several ways:
four equal-sized groups in each of
distance to nearest filament $D_F$,
stellar mass, and $u-r$ color;
and spiral, elliptical, and edge-on galaxies.
We do not find significant alignments
for any of the groups, nor do we find
significant linear trends with any of these properties.

We also measure alignments with the
Bisous model filaments of \citet{temp+14},
and find similar results (Table~\ref{tab:alignments_tempel}).
While the overall alignments are
stronger for the Bisous filaments (1.16$\sigma$ versus 0.61$\sigma$),
in neither case are they statistically
significant, and we do not find statistically
significant alignments with any subsample
in mass, color, distance to filament, or morphological
type for the \citet{temp+14} filaments.
The similar alignment results with the two
filament finders, despite the drastic methodological
differences between the Bisous model
and the \textit{Cosmic Web Reconstruction}
filaments, bolster our conclusion that the
MaNGA galaxies lack significant alignments
with filaments.

\begin{figure}[h]
\hspace{-20pt}
\psfig{file=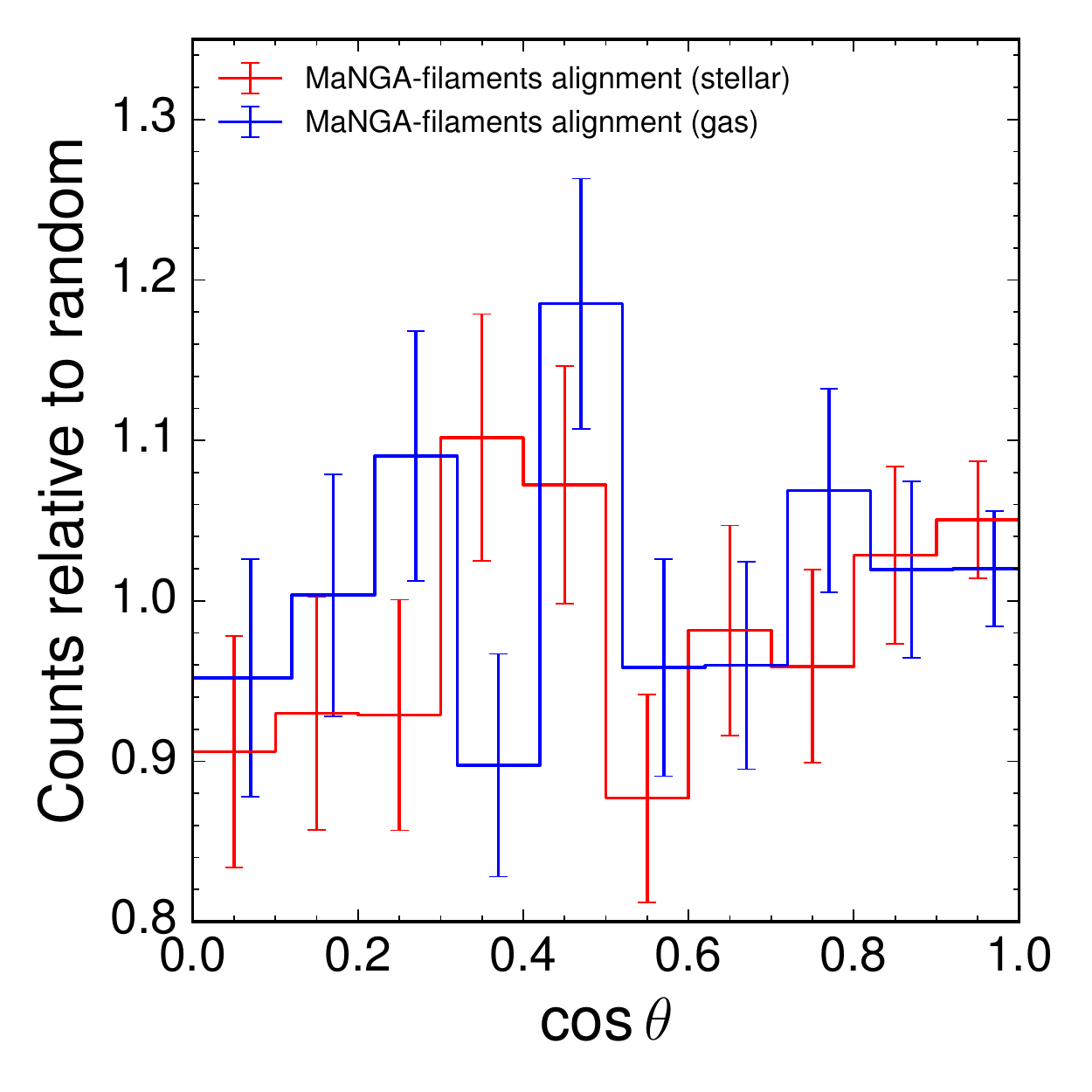,width=9.5 cm,clip=} 
\caption[]{\small Distribution of $\cos{\theta}$ (angle between galaxy spin and \textit{Cosmic Web Reconstruction} filament direction; red for stellar spins and blue for
emission line spins) compared to random alignments.
Each histogram is divided by the expectation
for random alignments in that bin.
Error bars are computed from Poisson statistics and the gas-filament alignment histogram is offset for clarity.
}
\label{fig:dot_product_histogram}
\end{figure}

Figure~\ref{fig:dot_product_histogram} shows that
the distribution of $\cos{\theta}$ is fully consistent
with random alignments.
The scatter in $\cos{\theta}$ is dominated
by intrinsic scatter in the alignments between galaxy spins and filaments,
rather than measurement error from either the galaxy spins or the filament directions.  By measuring
the total scatter in the galaxy-filament
alignments
and subtracting the contribution
from measurement error in quadrature,
we can estimate the intrinsic
scatter in alignments between
galaxy spins and filaments.
We estimate the contribution from measurement error
by creating 50,000 realizations of the alignment dataset
in which each filament or position angle is drawn from a Gaussian
with standard deviation given by the reported measurement error.  We find
that the standard deviation of the resulting mean
dot product 
(i.e.\ the scatter from measurement error) is 0.0044.  The total standard error of 0.0075 is slightly higher
than the standard error expected if the galaxies and filaments
were entirely randomly aligned, 0.0074.
Since the standard error cannot extend higher than $\sim 0.0074$,
at this point the quadrature sum of the intrinsic scatter
and measurement error may exceed the total scatter,
and thus we can only place a lower 
bound on the intrinsic scatter, $\sigma_i \geq 0.0061$.

\begin{deluxetable*}{lccccc}
\tablecolumns{6}
\tablecaption{\label{tab:alignments_tempel} Bisous filament alignments for different subsamples}
\tablehead{
Sample & N & $\langle \cos{\theta} \rangle$ & SE  & Shuffle mean & $\sigma$ from shuffle
}
\startdata
All & 2546 & 0.6462 & 0.0079 & 0.6370 &  1.16 \\
$D_F < 0.3$ Mpc & 635 & 0.6265 & 0.0178 & 0.6357 & -0.57 \\
$0.3 < D_F < 1.0$  Mpc & 637 & 0.6360 & 0.0151 &  0.6375 & -0.04 \\
$1.0 < D_F < 1.8$ Mpc  & 636 & 0.6493 & 0.0152 & 0.6402 & 0.83 \\
$D_F > 1.8$ Mpc & 638 & 0.6716 & 0.0150 & 0.6363 & 2.32  \\
$M_{\star} < 9.89$ & 636 & 0.6377 & 0.0137 & 0.6369 & 0.08  \\
$9.89 < M_{\star} < 10.38$ & 637 & 0.6619 & 0.0123 & 0.6371 & 2.05  \\
$10.38 < M_{\star} < 10.82$ & 636 & 0.6487 & 0.0130 & 0.6388 & 0.94  \\
$M_{\star} > 10.82$ &  637 & 0.6259 & 0.0160 & 0.6363 & -0.67  \\
$u-r < 1.65$ & 636  & 0.6369 & 0.0151 & 0.6367 & 0.02 \\
$1.65 < u-r < 2.06$& 637 & 0.6679 & 0.0146 & 0.6370 & 2.15 \\
$2.06 < u-r < 2.33$ & 636 & 0.6343 & 0.0142 & 0.6393 & -0.16 \\
$u-r > 2.33$ & 637 & 0.6498 & 0.0159 & 0.6343 & 0.83  \\
elliptical & 1030 & 0.6489 & 0.0122 & 0.6368 & 1.01  \\
spiral & 667 & 0.6558 & 0.0147 & 0.6361 & 1.30 \\
edge-on & 338 & 0.6262 & 0.0211 & 0.6396 & -0.50  \\
\enddata
\tablecomments{MaNGA spin-Bisous filament alignments for the entire sample and sub-samples split by distance
to filament ($D_F$), stellar mass (in units of $\log{ M_{\odot}}$), $u-r$ color, and morphology.  $\langle \cos{\theta} \rangle$ is the mean dot product
between the unit spin vector and the unit filament vector.  SE is the standard error of the mean, calculated using 50000 bootstrap resamples of the data.  We measure the expectation for random alignments
using 50000 shuffles of the data, and compute $\sigma$, the deviation between the data and the randoms in units of the standard error.}
\end{deluxetable*}

\section{Mass-dependence of spin-filament alignments}
\label{sec:massdependence}


Previous work has found that galaxy spin-filament alignments
in $N$-body and hydrodynamical simulations are mass dependent,
with lower mass galaxies showing alignment and higher mass
galaxies showing anti-alignment \citep{2005ApJ...627..647B,ac+07,hahn+07_evolution,codis+12,trow+13,ac14,dub+14,codis_IA}.  As a result of this mass dependence, it is possible
that a significant alignment signal could be concealed 
by opposing contributions from high and low mass galaxies.  Therefore,
we study the mass dependence of the alignment signal
and compare it to mass-dependent alignments in the MassiveBlack-II
and Illustris-1 simulations.  We attempt to mimic the construction
of the spin and filament catalogs as closely as possible to present a fair and quantitative
comparison between data and simulations.

We separate galaxies in the data and simulations
into five bins of $\Delta \log{M_{\star}} = 0.5$, with the lower limits of each bin ranging from $10^{9.6} M_{\odot}$ to $10^{11} M_{\odot}$ (the lowest bin has $\Delta \log{M_{\star}} = 0.4$).
We ignore galaxies less massive than $10^{9.6} M_{\odot}$ because MaNGA is incomplete below this mass, yielding a sample of 2551 MaNGA galaxies.  While the individual stellar masses have relatively large uncertainties
\citep[$0.2-0.3$ decades; ][]{blanton+roweis:2007,conroy:2013}, each bin in stellar mass has $>100$ galaxies
and thus the mass uncertainties are much smaller than the bin sizes.

The redshift distribution of each stellar mass bin is quite different due
to the strong correlation between redshift and stellar mass
in the MaNGA sample (Figure~\ref{fig:redshift_mass_dist}).
Furthermore, the number density of galaxies in the SDSS Main
Galaxy Sample is a strong function of redshift, and thus
the fidelity of recovery of the filaments will
be better at low redshift than at high redshift. These effects may introduce a spurious mass-dependence
into the alignment signal.  The hydrodynamical simulation
boxes are only 100 \hMpc{}, so we cannot  create
a lightcone mocking the SDSS Main Galaxy Sample.
Instead, we create 26 different realizations of the filament
catalogs with filaments found using different subhalo densities (i.e.\ representing different galaxy densities),
corresponding to the redshift slices of the filament catalog,
as described in Section~\ref{sec:mbii}.
In each realization, we find filaments in two dimensions
as in the data.
We assign each galaxy to one of the 26 different filament catalogs
by drawing from the redshift distribution of the MaNGA galaxies
at a given mass, weighted by the MaNGA
volume weights.  In this way, we assign each galaxy in the simulation
to a unique filament, and measure the two-dimensional spin-filament alignment
in the same manner as in the data.
We estimate error bars using the standard error of the mean of each bin,
and average over 100 random draws from the mass-redshift distribution.
We assess the discrepancy between data and simulation using $\chi^2$,
with errors given by the quadrature sum of the errorbars on the data
and errorbars on the simulation.

\begin{figure}
\hspace{-20pt}
\includegraphics[width=0.55\textwidth,clip=true,trim=0 0 0 0]{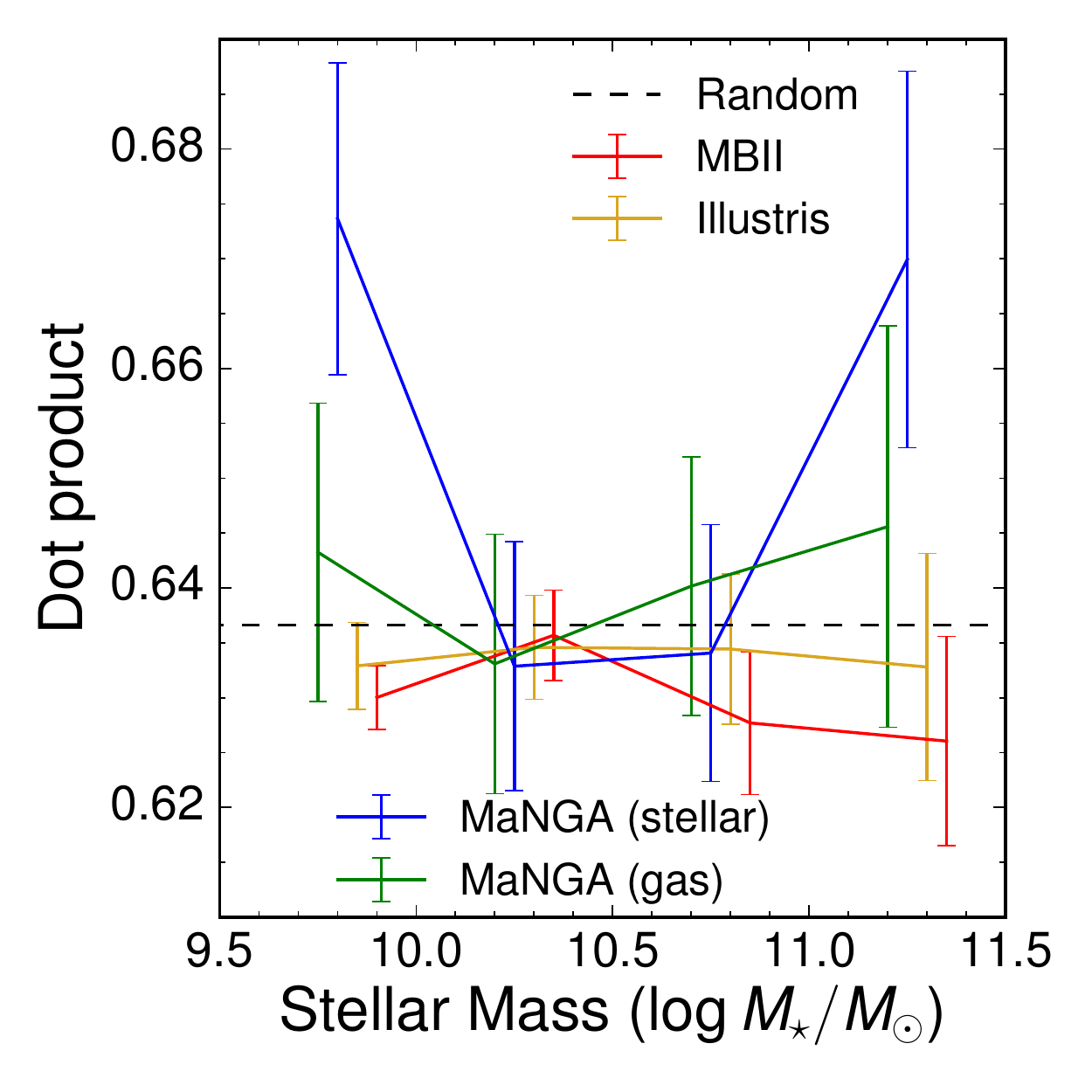}
\caption[]{\small Mass-dependence of MaNGA spin-filament alignment,
comparing MaNGA alignments using both stellar (blue) and emission line (green) spins
to alignments in the MassiveBlack-II (red) and Illustris (orange) simulations.
In each bin, points from different
samples are offset for clarity.}
\label{fig:alignments_by_sample}
\end{figure}

This methodology yields different spin-filament alignments from the standard picture, with weak anti-alignments seen at all masses,
rather than a transition from alignments at low mass to anti-alignments at high mass.
This discrepancy arises from the enforced degeneracy between mass and redshift: at 
high mass the sample
is dominated by high redshift galaxies, which are associated with more poorly measured
filaments due to the lower number density in the Main Galaxy Sample at higher redshift.  
While high mass galaxies show stronger anti-alignments than low mass galaxies at fixed
redshift,  the strong
anti-alignment at high masses is weakened by the degeneracy
between mass and redshift.
The difference between the simulation curves in Figure~\ref{fig:alignments_by_sample}
and the standard picture highlights the importance of constructing a simulation
sample that closely mimics the methodology of the data.

For the fiducial case, we find modest tension between
the mass-dependence of alignments in the MaNGA sample of 2551 galaxies
and the mass-dependence in the hydrodynamical simulations, with
$\chi^2 = 14.26$ over 4 degrees of freedom ($p = 0.0065$, equivalent
to 2.7$\sigma$) for MassiveBlack-II
and $\chi^2 = 11.09$/4 dof ($p = 0.026$, equivalent to 2.2$\sigma$) for Illustris. We find similar 2-3$\sigma$ tensions
when using different bins, and in fact find a higher $\chi^2 = 15.52$
when using a stricter cut of $\sigma_{\textrm{PA}} < 3^{\circ}$, indicating that the tension is not an
artifact of the binning scheme and cuts used.
However, this tension is clearly absent in the mass-dependence
of alignments for H$\alpha$ emission line spins, for which we find
$\chi^2 = 2.59$ over 4 degrees of freedom between the data
and MassiveBlack-II.

We confirm that the $\chi^2$ test is appropriate for this comparison: the mean dot product in each stellar mass bin
is normally distributed, and the covariance between neighboring bins is small compared
to the variance of each bin.  Using 50,000 bootstrap resamples, we confirm that the distribution of the mean
dot product in each mass bin is normally distributed, even in cases where there are only $\sim 50$ galaxies
in the smallest (most-massive) bin. We estimate the covariance by resampling the galaxies in 100 deg$^2$
blocks rather than re-sampling galaxy by galaxy in order to preserve the source of the covariance,
correlations between neighboring galaxy spin-filament dot products arising from galaxy spin correlations,
which drop rapidly over $\sim 10$ \hMpc{} \citep{pen:2000,lee11}.
We find that computing the $\chi^2$ with this covariance matrix rather than assuming a diagonal
 covariance matrix makes little difference, and that the 
resulting covariance matrices are relatively robust to changes in the size of the 
resampling blocks.  Given that MaNGA galaxies are widely distributed over the sky,
with the average pair separation greater than the $\sim 10$ \hMpc{} spin correlation length (Figure~\ref{fig:filaments_galaxies}), we expect the covariance matrix
to be nearly diagonal.

The sample of MaNGA galaxies with well measured spins is not complete: in fact, the completeness varies as a function of mass,
with low and high mass galaxies having relatively low completeness of well-measured stellar spins, whereas intermediate-mass
galaxies are quite complete (Table~\ref{tab:completeness}).  This incompleteness preferentially selects galaxies with higher specific angular momentum $j$, for which it is easier to measure a spin direction.  This could
possibly bias the mass-dependence of spin-filament alignments, if high $j$ galaxies have different alignments
than low $j$ galaxies.  We attempt to estimate the bias introduced by this incompleteness by removing low $j$ subhalos
in the mass bins in the simulation to match the incompleteness of stellar spins in MaNGA.  This is a conservative procedure,
as incompleteness is likely also caused by low-S/N stellar continuum and plane-of-sky inclinations, which are not related to galaxy-filament alignment stength.  Nevertheless,
removing low-$j$ subhalos has an extremely modest effect
on alignments in the simulations, 
changing the $\chi^2$ between data and Illustris from 11.09 to 12.57.

These tests suggest that the comparison between data and simulations presented above is not impacted by covariance between the stellar mass bins or incompleteness in the spin measurements.  Therefore the discrepancy between the spin-filament alignments in data and simulations remains unresolved.

\begin{deluxetable}{lcc}
\tablecolumns{3}
\tablecaption{\label{tab:completeness}}
\tablehead{
Bin & Stellar & Emission line \\
 & completeness & completeness}
\startdata
$ 9.6 < \log{h^{-1} M_{\odot}} < 10$ & 83.7\% & 91.3\% \\
$ 10 < \log{h^{-1} M_{\odot}} < 10.5$ & 97.1\% & 91.3\% \\
$ 10.5 < \log{h^{-1} M_{\odot}} < 11$ & 95.6\% & 90.0\% \\
$ \log{h^{-1} M_{\odot}} > 11$ & 76.2\% & 76.2\% \\
\enddata
\tablecomments{Fraction of MaNGA galaxies with well measured spins (error $< 5^{\circ}$)
in different mass bins.}
\end{deluxetable}

\subsection{3D alignments in simulations}
\label{sec:3d}

While the mass-dependent alignment signal in data is quite modest,
more significant differences become apparent if we instead
use simulated filaments with no regard to observational constraints, i.e.\ filaments in three dimensions
and filaments measured using all subhalos, rather than only
using massive observable galaxies.  This allows us to detect
galaxy-filament alignments at much higher significance.

\begin{figure*}
\includegraphics[width=\textwidth,clip=true,trim=0 100 0 120]{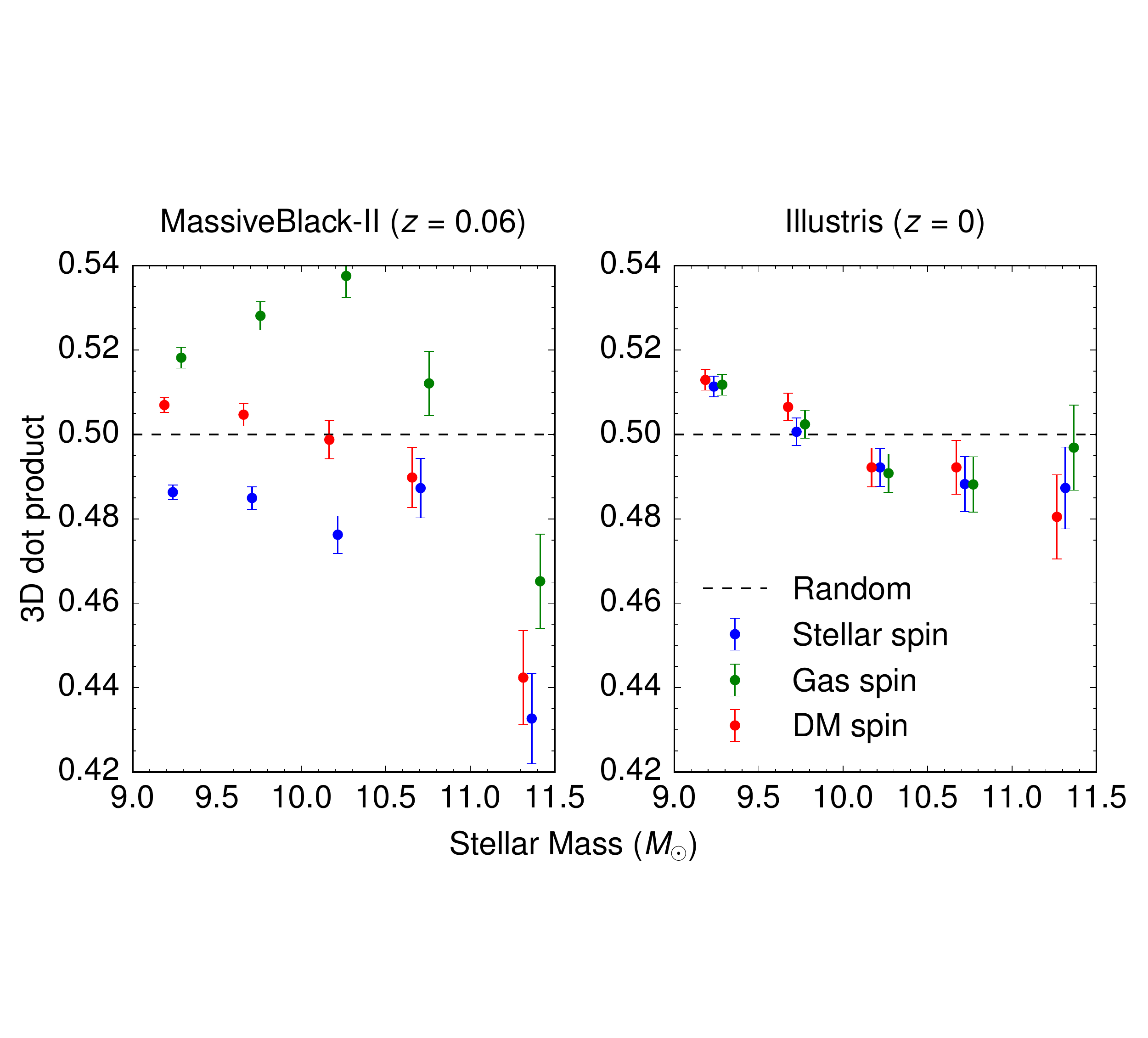}
\caption[]{\small 3D alignments between subhalo spins and ``ideal''
filaments measured using the 500,000 most-massive subhalos in Illustris ($z = 0$)
and MassiveBlack-II ($z = 0.06$).  In both panels, the black dotted line
is the 3D dot product expected for random alignments,
the red (blue; green) points are alignments between dark matter (stellar; gas) spins and filaments. Points are offset in mass for clarity.
\vspace{10pt}
}
\label{fig:illustris_vs_mbii}
\end{figure*}

Both MassiveBlack-II and Illustris show similar mass-dependence of the alignments between
dark matter spins and filaments (Figure~\ref{fig:illustris_vs_mbii}).
This transition from aligned at low masses to anti-aligned at higher masses is consistent
with previous findings, mostly from dark-matter-only simulations
\citep{2005ApJ...627..647B,ac+07,hahn+07_evolution,pich+11,codis+12,trow+13,ac14,dub+14,codis_IA,wang18b}.

In contrast, Illustris and MassiveBlack-II paint opposing pictures of the mass dependence of stellar spin-filament
alignments.
In Illustris the mass-dependence of the stellar spin alignments is quite similar to the mass-dependence
of the dark matter spin alignments, while in MassiveBlack-II the stellar spin alignments show a qualitatively different behavior
than the dark matter spin alignments, remaining anti-aligned even at the lowest masses (Figure~\ref{fig:illustris_vs_mbii}).  The
$z \sim 0$ results in Illustris are consistent with the findings of \citet{dub+14} in the Horizon-AGN hydrodynamic simulation
at $z = 1.8$, who measured alignments between filaments and stellar angular momentum and found a transition
from alignment to anti-alignment at $M_{\star} \sim 10^{10.5}$ $M_{\odot}$.
Additionally, we find very similar
mass-dependent alignments to \citet{wang18b}, who also use
Illustris to measure alignments between galaxies and the third eigenvector
of the deformation tensor, which defines the filament direction within the Zel'dovich approximation.

The mass-dependent trends in gas spin
alignments are even more divergent between Illustris
and MassiveBlack-II.  Gas-filament alignments
in Illustris are quite similar to star-filament
and dark matter-filament alignments, but gas spins in MassiveBlack-II remain aligned with filaments until
$M_{\star} \sim 10^{11}$ $M_{\odot}$, and the alignments
are considerably stronger than low-mass dark matter-filament alignments.  While gas spins in MassiveBlack-II are only measured for a subset of
galaxies with $> 1000$ gas particles, this subset has very similar stellar-filament and dark matter-filament alignments as the entire sample, implying
that the gas-filament alignments are not significantly impacted by selection bias.

Taken together, these results suggest that
while the ``transition mass'' picture presented in previous work
\citep[e.g.][]{codis+12}
remains valid for dark matter spins, its validity for stellar and gas spins of galaxies is questionable and apparently dependent on subgrid physics and feedback models. 

In conclusion, we find modest tension between the mass dependence of galaxy-filament
alignments in MaNGA and in the MassiveBlack-II and Illustris simulations.
The tension is present if MaNGA spins
are estimated using stellar continuum
velocities,
although 
it disappears if we use MaNGA spins measured from the H$\alpha$ emission
line.
While we find minimal differences in alignments between MassiveBlack-II
and Illustris using a sample of simulated galaxies and filaments
selected to mimic the MaNGA and SDSS galaxy samples, an ideal measurement
using filaments constructed from all subhalos in the simulations
reveals a significant difference in the behavior of spin alignments
in Illustris and MassiveBlack-II at low masses. While both simulations
find that dark matter spins are aligned
with filaments at low mass, in agreement
with previous results
from $N$-body simulations,
Illustris finds stellar spin-filament alignment at low mass, while MassiveBlack-II
finds stellar spin-filament anti-alignment.

\section{Discussion}
\label{sec:conclusions}

We present the first measurement of alignments between filaments and galaxy spins as measured from integral-field kinematics.
We find  no significant detection of galaxy spin alignments with filaments.
We find that the mass dependence of spin-filament alignments
from MaNGA are in 2-3$\sigma$ tension with spin-filament alignments
from the MassiveBlack-II and Illustris simulations, although the
tension disappears if we instead use galaxy spins measured
from the H$\alpha$ emission line.
While
the predictions of MassiveBlack-II and Illustris are essentially
identical if we use a ``mock-observational'' sample, 
three-dimensional filaments measured using all subhalos
in the simulation reveal significant differences in alignment
behavior at low masses, suggesting that the ``transition-mass''
picture described in previous works is dependent on details
of feedback and subgrid physics.
 
 Previous studies have measured galaxy spin-filament alignments using galaxy shape as a proxy for galaxy
 spin \citep[]{temp13,tl13,pah+16,chen+18}.  These studies find
 a weak dichotomy between spiral and elliptical galaxies, with
 spirals aligned and ellipticals anti-aligned with filaments.
 We do not find evidence for this dichotomy, but our
 error bars are larger than in previous studies and our results are consistent with them.

 The results in this work are limited by the relatively small sample size of $\sim 2600$ MaNGA galaxies
 with well-measured spins and sufficient proximity to ``Cosmic Web Reconstruction'' filaments.
 The error bars on this measurement are dominated by intrinsic scatter rather than measurement
 error on the spins or the filaments, suggesting that acquiring larger samples of galaxy spins is the
 most effective way to achieve a more precise measurement.
 The full MaNGA sample will provide integral-field-unit spectroscopy for 10,000 galaxies,
 roughly doubling the sample with sufficient spin measurements. 
 This represents a significant step forward, but even larger samples are needed to 
 distinguish the alignment models of different simulations at high significance.
 The proposed Hector survey on the Anglo-Australian Telescope could deliver integral-field
 spectroscopy for up to 100,000 galaxies over the next decade \citep{bryant:2016}, offering
 an unparalleled ability to learn about the relationship between galaxy spin
 and large-scale structure and the acquisition of galaxies' angular momentum.

\acknowledgements{
We thank Simone Ferraro, Joanne Cohn and Shy Genel for helpful comments
and discussions.

This research made use of Marvin, a core Python package and web framework for MaNGA data, developed by Brian Cherinka, Jos{\'e} S{\'a}nchez-Gallego, Brett Andrews, and Joel Brownstein \citep{brian_cherinka_2018_1146705}.

Calculations presented in this paper used resources of the National Energy Research Scientific Computing Center (NERSC), which is supported by the Office of Science of the U.S.~Department of Energy under Contract No.~DE-AC02-05CH11231.

Funding for the Sloan Digital Sky Survey IV has been provided by the Alfred P. Sloan Foundation, the U.S. Department of Energy Office of Science, and the Participating Institutions. SDSS acknowledges support and resources from the Center for High-Performance Computing at the University of Utah. The SDSS web site is \url{www.sdss.org}.

SDSS is managed by the Astrophysical Research Consortium for the Participating Institutions of the SDSS Collaboration including the Brazilian Participation Group, the Carnegie Institution for Science, Carnegie Mellon University, the Chilean Participation Group, the French Participation Group, Harvard-Smithsonian Center for Astrophysics, Instituto de Astrof{\'i}sica de Canarias, The Johns Hopkins University, Kavli Institute for the Physics and Mathematics of the Universe (IPMU) / University of Tokyo, Lawrence Berkeley National Laboratory, Leibniz Institut f{\"u}r Astrophysik Potsdam (AIP), Max-Planck-Institut f{\"u}r Astronomie (MPIA Heidelberg), Max-Planck-Institut f{\"u}r Astrophysik (MPA Garching), Max-Planck-Institut f{\"u}r Extraterrestrische Physik (MPE), National Astronomical Observatories of China, New Mexico State University, New York University, University of Notre Dame, Observat{\'o}rio Nacional / MCTI, The Ohio State University, Pennsylvania State University, Shanghai Astronomical Observatory, United Kingdom Participation Group, Universidad Nacional Aut{\'o}noma de M{\'e}xico, University of Arizona, University of Colorado Boulder, University of Oxford, University of Portsmouth, University of Utah, University of Virginia, University of Washington, University of Wisconsin, Vanderbilt University, and Yale University.

}

\appendix

\section{Galaxy spin fitting}
\label{sec:spin_details}

We use the \textsc{FIT\_KINEMATIC\_PA} routine \citep{kraj+06} to determine the kinematic position angle for each galaxy
from the stellar velocity maps, using velocities from the unbinned spaxels.
We remove low quality or potentially problematic data by masking spaxels with $r$-band SNR $< 5$,  
spaxels with the \textsc{DONOTUSE} or \textsc{UNRELIABLE} bitmasks (Westfall et al., in prep), $|v| > 350$ km s$^{-1}$,
$\sigma_v > 10^3$ km s$^{-1}$, or a velocity that is more than a 5-$\sigma$ outlier (i.e. $| v | > 5$ times the standard deviation of
$v$).
We also mask all contiguous regions with SNR $> 5$ that are disconnected from the central part of the galaxy in order to eliminate faint
companion galaxies within the IFU.  To avoid giving a large weight to any one spaxel,
we set the minimum velocity error to 2 km s$^{-1}$ \citep{pineda+2017}.
From visual inspection of the fits,
we find that these settings
give the best performance.
We recenter each galaxy about the unweighted centroid of its unmasked region, since the center of rotation in some galaxies is offset from 
the center of the IFU. 
Finally, we perform each fit in curved-sky coordinates ($\alpha \cos{\delta}$, $\delta$) and convert the resulting position angles
to an equirectangular projection for consistency with the filament catalog.  We use this method as it fits the position
angles in a physical coordinate system.

\textsc{FIT\_KINEMATIC\_PA} fits a bi-antisymmetric model to a velocity map.  For a specified rotation of the $xy$ coordinates
relative to the native ($\alpha \cos{\delta}$,$\delta$) coordinates (i.e.\ position angle), 
the bi-antisymmetric model at ($x$,$y$) is the average of the velocity at ($\pm x$, $\pm y$), linearly interpolating between
neighboring points if need be.  The best-fit position angle minimizes $\chi^2$ computed from the data, the bi-antisymmetric model and the MaNGA
velocity errors.
We initially loop over all PAs between 0$^{\circ}$ and 180$^{\circ}$ to ensure that
we are near the global minimum, then use Nelder-Mead minimization to find the global minimum $\chi^2$. 

To estimate the error on the position angle, we create 100 realizations of the velocity map, drawing the velocity in each spaxel
from a Gaussian centered at the measured velocity, with standard deviation equal to the velocity error, and assuming
no covariance between neighboring spaxels.  We apply the same $\chi^2$ minimization process to each of the 100 realizations, again
using the MaNGA velocity errors and the bi-antisymmetric model from \textsc{FIT\_KINEMATIC\_PA}.
We define the position angle as the mean of the ensuing 100 position angles $\theta_i$
and the
position angle error as the standard deviation of the 100 position angles.
We use the circular mean of headless (i.e.\ spin-2) vectors $\mu_{180}$:
\begin{equation}
\mu_{180}(\vec{\theta}) = \frac{1}{2} \arctan{\frac{\sum{\sin{2\theta_i}}}{\sum{\cos{2\theta_i}}}}
\label{eqn:mean180}
\end{equation}
The standard deviation is adjusted similarly:
\begin{equation}
\sigma_{180}(\vec{\theta}) = \sqrt{\frac{1}{N}\sum\textrm{min}^2(\theta_i - \mu_{180}(\vec{\theta}),180-\theta_i + \mu_{180}(\vec{\theta}))}
\label{eqn:std180}
\end{equation}

While this is only an approximate estimate of the position angle error, and may in particular underestimate
the error due to nonzero covariance between neighboring spaxels, the position angle errors are not an important
contributor to the total error budget on the alignment measurements: they
are dominated by the errors
on the filament angles (Figure~\ref{fig:fils_vs_pa_err}) and as we show in Section~\ref{sec:methods}, the measurement errors
are dominated by the intrinsic scatter in galaxy-filament alignments anyway.

\section{3D vs 2D alignment measurements in simulations}
\label{sec:3d_vs_2d}

We use measurements of galaxy-filament alignments in the MassiveBlack-II
simulation to determine how much signal is lost using two-dimensional
rather than three-dimensional measurements of the filaments.
We generate three-dimensional filaments using the redshift-space positions
of the top 500,000 subhalos
in MBII by total mass ($\log{M_h/M_{\odot}} > 9.74$) and applying the Cosmic Web Reconstruction
algorithm with a smoothing bandwidth of 1 $h^{-1}$ Mpc.  For the two-dimensional
sample, we use the same subhalos and bandwidth, but as in Section~\ref{sec:mbii},
we split the box into 7 slices along the $z$ direction ($\Delta z= 0.005 = 20$ Mpc $\sim$ 14 $h^{-1}$ Mpc)
and separately find two-dimensional filaments in each slice.
In both cases, the number density of subhalos is much greater than
the number density achievable in the Main Galaxy Sample; we use these large
samples in order to make a high signal-to-noise measurement of three-dimensional
and two-dimensional alignment, and assume that the reduction
in signal from three-dimensional to two-dimensional will be similar
for the realistic lower number-density samples.

We find that the mean dot product between the 3D filaments and the subhalo
stellar spins is $0.4882 \pm 0.00136$ compared to an expectation of 0.5
for random alignments for an alignment strength signal-to-noise of 8.68.  For the two-dimensional filaments, we find
a mean dot product of $0.6293 \pm 0.00145$ compared to 0.6366 for random alignments,
yielding an alignment strength signal-to-noise of 5.03.
We also measure alignments between subhalo spins and ``ideal'' two-dimensional
filaments, which are the projection of the 3D filaments onto the $xy$ plane;
here we find a mean dot product of $0.6267 \pm 0.00146$ and signal-to-noise 6.81.
This indicates that most of the reduction in the $\Delta z = 0.005$ case comes from
loss of filament information in the $z$ direction and not from the finite  $ \Delta z$ of the slices.
Therefore, we estimate that using two-dimensional filaments in slices of $\Delta z= 0.005$ reduces the alignment
signal strength by 40\%.  However, it is unlikely that we could realize a 40\% improvement in signal by using
three-dimensional filaments, since the line-of-sight component of the galaxy spin vector is significantly harder
to measure than the transverse component, reducing the signal gain from three-dimensional
filaments \citep{krolewski:2017}.

\bibliographystyle{yahapj}
\bibliography{citations.bib}

\end{document}